\documentclass{webofc}
\usepackage[varg]{txfonts}   
\begin{document}
\title{Photometric Redshift Analysis using Supervised Learning Algorithms and Deep Learning}
%
%

\author{\firstname{} \lastname{Chong} \firstname{De Wei, Kenny} \inst{1}\fnsep\thanks{\email{kenny.chong@u.nus.edu}} \and
        \firstname{Abel} \lastname{Yang}\inst{1}\fnsep\thanks{\email{phyyja@nus.edu.sg}}
}

\institute{Department of Physics, National University of Singapore}

\abstract{%
We present a catalogue of galaxy photometric redshifts for the Sloan Digital Sky Survey (SDSS) Data Release 12. We use various supervised learning algorithms to calculate redshifts using photometric attributes on a spectroscopic training set. Two training sets are analysed in this paper. The first training set consists of 995,498 galaxies with redshifts up to $z \approx 0.8$. On the first training set, we achieve a cost function of 0.00501 and a root mean squared error value of 0.0707 using the XGBoost algorithm. We achieved an outlier rate of 2.1\% and 86.81\%, 95.83\%, 97.90\% of our data points lie within one, two, and three standard deviation of the mean respectively. The second training set consists of 163,140 galaxies with redshifts up to $z\approx0.2$ and is merged with the Galaxy Zoo 2 full catalog. We also experimented on convolutional neural networks to predict five morphological features (Smooth, Features/Disk, Star, Edge-on, Spiral). We achieve a root mean squared error of 0.117 when validated against an unseen dataset with over 200 epochs. Morphological features from the Galaxy Zoo, trained with photometric features are found to consistently improve the accuracy of photometric redshifts.}
\maketitle
\section{Overview and Methodology}
Redshifts of celestial objects have been a vital component in the field of astronomy and we use them to measure various attributes such as the rotation of the galaxy and the distance from us. Traditionally, they have been measured by spectroscopy. While spectroscopy is effective in determining redshifts of galaxies, it is time consuming and expensive and therefore not scalable to map a spectroscopic redshift for every celestial object. Spectroscopy data is limited in the SDSS database with only 0.36\% mapped \cite{2011AJ....142...72E}. There are two methods in the literature currently deployed to determine photometric redshifts: template-fitting and the statistical/machine learning method. We will focus on the machine learning method and also use deep learning methods to increase predictive power.

The SDSS is a major multi-filter imaging and spectroscopic redshift survey using a 2.5m wide angle optical telescope. We queried out a data size of 995,498 galaxies with redshifts up to $z \approx 0.8$ for our training and validation set. We first split the data points and photometric features using the random subspace method. For each subset, we train them on an individual tree and use standard deviation reduction to determine the quality of every split. The process is repeated when building the tree down to the leaf node. We use this method to find the more important features.

At the same time, understanding the physics behind the various photometric features allow us to feature engineer attributes that perform stronger than the native features. In another paper investigating photometric redshifts \cite{2008ApJ...674..768O}, concentration indices are found to break the degeneracy between colour band and photometric redshifts. Indeed, concentration indices contain some information on morphological classification and concentration indices separate galaxies into early and late types \cite{2003AJ....125.1682N}, with the early type galaxies having lower values of the concentration indices. This motivates the need to include the concentration indices into the neural network.

Since redshifts are determined by looking at transition lines and these lines form very sharp peaks at certain wavelengths, we are able to extract some photometric information on these peaks by taking the difference between the colour bands. The colour bands and colour indexes interact with each other and we use the multilayer perceptron to encourage these interactions.

For each neural network, we perform a 5-fold validation and the same neural network is trained 5 times. The average results of the cost function are then recorded.
\section{Results}
\begin{table}
\centering
\begin{tabular}{ll}
\textbf{Model Features} & \textbf{Cost Function}\\
\hline
modelmag color band& 0.0013276 \\
(u, g, r, i, z) & \\
dered color band & 0.0013093 \\
(u, g, r, i, z) & \\
dered color indexes & 0.0014040\\
(u-g, g-r, r-i, i-z) \\
dered color band + dered color indexes & 0.0012729 \\
dered color band + dered color indexes + concentration indices & 0.0012559\\
\hline
\end{tabular}
\caption{Cost Function of Neural Networks with various features}
\label{tab-1}
\end{table}

As shown in Table \ref{tab-1}, the best model consists of 14 features with a cost function of 0.0012559 with the second training set. The main dataset is then trained with the best model features across 4 algorithms (Linear Regression, Random Forests, XGBoost and Neural Networks [MLP]) and we achieve a cost function of 0.00501 using the XGBoost model \cite{Chen:2016:XST:2939672.2939785}.

\begin{figure}
\centering
\includegraphics[width=\textwidth,clip]{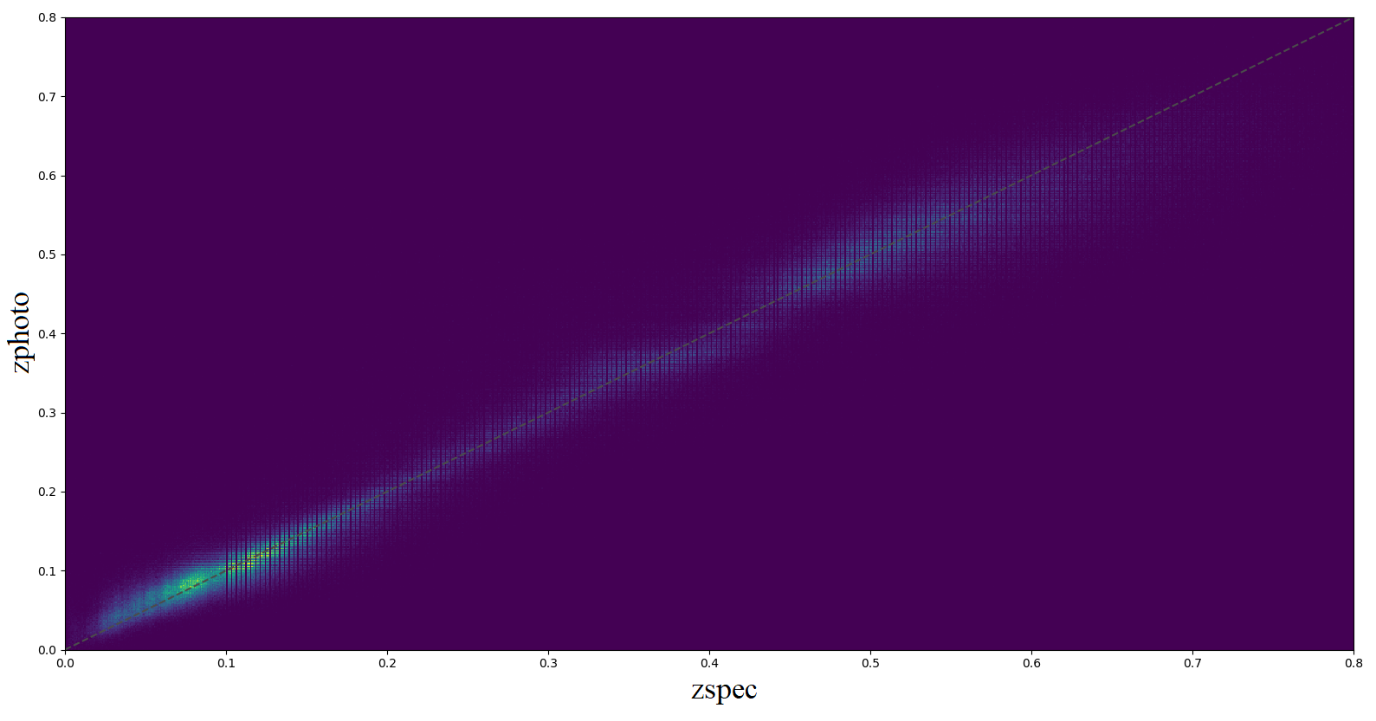}
\caption{XGBoost model of first training set}
\label{fig-1}       
\end{figure}

In Figure \ref{fig-1}, we plot photometric determined redshifts versus spectroscopic redshifts in a gaussian density plot and the average redshift bias is determined to be $2.59 \times 10^{-3}$ with a standard deviation of 0.0559. As seen in Figure \ref{fig-1}, photometric redshift precision is found to decrease as the wavelength increases. The cost function is found to reduce to 0.00183 from 0.00501 for $r < 20$ spectroscopic objects whereas the cost function increase to 0.0106 for $r > 20$ spectroscopic objects.

Outliers are defined to be larger than three standard deviations of the mean. We achieved an outlier rate of 2.10\% and calculated 86.81\%, 95.83\%, 97.90\% of our data points lie within one, two, and three standard deviation of the mean respectively.

\section{Creating Scalable Features beyond SDSS photometric data}
\subsection{Performance of Galaxy Zoo 2 Features}
One major limitation with current photometric features is the degeneracy between spectroscopic determined redshifts and 5 colour SDSS photometric data. There is a considerable number of celestial objects with the same colour band range yet they give rise to different redshift values. No matter what supervised learning algorithm you use, it won't be able to differentiate these galaxies apart.

We attempt to use morphological features to break the degeneracy between redshift and colour band. The Galaxy Zoo \cite{2013MNRAS.435.2835W} is a crowd sourcing project where online citizens can contribute by classifying galaxies through a decision tree process into various categories. With a sufficient sample size, the end result takes on a value that is analogous to probability, that is if 60\% of online citizens voted a spiral pattern, the probability of a galaxy being a spiral is 0.6. Together with features from the SDSS, we merge the two databases together and deploy the same methodology to determine the performance of Galaxy Zoo features.
\begin{table}
\centering
\begin{tabular}{ll}
\textbf{Model} & \textbf{Cost Function}\\
\hline
Models without Galaxy Zoo 2 Features & 0.001256 \\
Models with Galaxy Zoo 2 Features & 0.001198 \\
\hline
\end{tabular}
\caption{Performance of Galaxy Zoo 2 Features}
\label{tab-2}
\end{table}

The selected Galaxy Zoo 2 features are the probability of galaxy being smooth, the probability of galaxy having features/disk, probability of galaxy, probability of galaxy being an edge-on galaxy and probability of galaxy having a spiral. Similarly, we train neural networks with the same subset of features in Table \ref{tab-1} together with Galaxy Zoo 2 features and determine their performance using the cost function. Neural networks with Galaxy Zoo 2 features are found to consistently outperform neural networks without Galaxy Zoo 2 features. This demonstrates these five Galaxy Zoo 2 features provided morphological information to break some degeneracy between redshift and information from the SDSS.

\subsection{Deep Learning}
The objective of this component is to build a model that predicts morphological shapes labelled by online citizens. We use a convolutional neural network architecture to train a multi-label image classifier. We down sample the input into 3 normalized channels (RGB) with $150\times150$ inputs. After several iterations, the final model consists of a total of 10 layers: 8 convolution layers, 4 max pooling layers, 2 full connected (with dropout) layers. In all convolution layers, we apply the `same' padding to retain the output shape of the matrix. A `ReLu' activation is applied, and the number of kernels used is 32, 32, 64, 128 in that order. Every two convolution layers is followed by a max pooling layer of size ($2\times2$). The output is then flattened into a vector and then fed into the fully connected layer with 1024 nodes, followed by an activation function. 50\% of the features are then dropped out. Similar to the MLP, the cost function used is the mean squared error and the `adam' optimizer \cite{2014arXiv1412.6980K} is used with learning rate = 0.001, beta1 = 0.9, beta2 = 0.999, fuzz factor = $10^{-8}$ and learning rate decay = 0.

We achieve a root mean squared error of 0.117 when validated against an unseen dataset with over 200 epochs. In addition, we can also map other morphological classifications into the model as well, and deploy transfer learning or use a more extensive architecture to predict the morphological features of any arbitrary galaxy to a higher accuracy. Deep learning algorithms allow us to scale morphological features without the need of manual labelling and we can use the morphological features from the galaxy zoo to determine photometric redshifts to a higher accuracy.


\begin{thebibliography}{6}

\bibitem{2011AJ....142...72E}
D.J. {Eisenstein}, D.H. {Weinberg}, E.~{Agol}, H.~{Aihara}, C.~{Allende
  Prieto}, S.F. {Anderson}, J.A. {Arns}, {\'E}.~{Aubourg}, S.~{Bailey},
  E.~{Balbinot} et~al., Astron. J. \textbf{142}, 72 (2011), \texttt{1101.1529}

\bibitem{2008ApJ...674..768O}
H.~{Oyaizu}, M.~{Lima}, C.E. {Cunha}, H.~{Lin}, J.~{Frieman}, E.S. {Sheldon},
  Astrophys. J. \textbf{674}, 768 (2008), \texttt{0708.0030}

\bibitem{2003AJ....125.1682N}
O.~{Nakamura}, M.~{Fukugita}, N.~{Yasuda}, J.~{Loveday}, J.~{Brinkmann}, D.P.
  {Schneider}, K.~{Shimasaku}, M.~{SubbaRao}, Astron. J. \textbf{125}, 1682
  (2003), \texttt{astro-ph/0212405}

\bibitem{Chen:2016:XST:2939672.2939785}
T.~Chen, C.~Guestrin, \emph{XGBoost: A Scalable Tree Boosting System}, in
  \emph{Proceedings of the 22Nd ACM SIGKDD International Conference on
  Knowledge Discovery and Data Mining} (ACM, New York, NY, USA, 2016), KDD '16,
  pp. 785--794, ISBN 978-1-4503-4232-2, \texttt{1603.02754}

\bibitem{2013MNRAS.435.2835W}
K.W. {Willett}, C.J. {Lintott}, S.P. {Bamford}, K.L. {Masters}, B.D. {Simmons},
  K.R.V. {Casteels}, E.M. {Edmondson}, L.F. {Fortson}, S.~{Kaviraj}, W.C.
  {Keel} et~al., MNRAS \textbf{435}, 2835 (2013), \texttt{1308.3496}

\bibitem{2014arXiv1412.6980K}
D.P. {Kingma}, J.~{Ba}, ArXiv e-prints  (2014), \texttt{1412.6980}

\end{thebibliography}

\end{document}